\documentclass[reprint,aps,prstper]{revtex4-2}

\usepackage{graphicx}
\usepackage{hyperref}
\usepackage{footnote}
\usepackage{amsmath,amssymb}
\usepackage{multirow}
\usepackage{tabularx}
\usepackage{array,ragged2e}
\newcolumntype{P}[1]{>{\RaggedRight\arraybackslash}p{#1}}
\usepackage[utf8]{inputenc}
\usepackage{enumitem}
\usepackage{longtable}
\usepackage{supertabular,booktabs}
\usepackage{pdfpages}

\makeatletter
\AtBeginDocument{\let\LS@rot\@undefined}
\makeatother

\begin{document}

\title{Education for expanding the quantum workforce: Student perceptions of the quantum industry in an upper-division physics capstone course}%

\author{Kristin A. Oliver$^{1,2}$}
\email[]{kristin.oliver@colorado.edu}
\author{Victoria Borish$^{1,2}$}
\author{Bethany R. Wilcox$^{1}$}
\author{H. J. Lewandowski$^{1,2}$}
\affiliation{$^{1}$Department of Physics, University of Colorado, Boulder, Colorado 80309, USA}
\affiliation{$^{2}$JILA, National Institute of Standards and Technology and University of Colorado, Boulder, Colorado 80309, USA}

\date{\today}

\begin{abstract}
As quantum technologies transition out of the research lab and into commercial applications, it becomes important to better prepare students to enter this new and evolving workforce. To work towards this goal of preparing physics students for a career in the quantum industry, a senior capstone course called ``Quantum Forge'' was created at the University of Colorado Boulder. This course aims to provide students a hands-on quantum experience and prepare them to enter the quantum workforce directly after their undergraduate studies. Some of the course's goals are to have students understand what comprises the quantum industry and have them feel confident they could enter the industry if desired.  To understand to what extent these goals are achieved, we followed the first cohort of Quantum Forge students through their year in the course in order to understand their perceptions of the quantum industry including what it is, whether they feel that they could be successful in it, and whether or not they want to participate in it. The results of this work can assist educators in optimizing the design of future quantum-industry-focused courses and programs to better prepare students to be a part of this burgeoning industry.  
\end{abstract}

\maketitle


\section{\label{sec:Intro}Introduction}
As the calls for advancing Quantum Information Science and Technology (QIST) in the United States have grown, especially with the passing of the 2018 National Quantum Initiative Act (NQI) ~\cite{noauthor_enabling_nodate}, the need for a quantum-ready workforce has increased. This was reinforced when the U.S. government announced that the ``shortage of talent [in quantum technologies] constrains progress,'' indicating that the development of new talent in the realm of quantum technologies is crucial ~\cite{subcommittee_on_quantum_information_science_committee_on_science_of_the_national_science__technology_council_quantum_2022}. 

In order to develop this new talent and, therefore, a strong quantum workforce, we must provide students the opportunity to learn the skills and competencies required to work in this industry. Additionally, we must also help students understand what the quantum industry is and whether or not they are interested in participating in it. This suggests that specific QIST education at various levels is 
vital to expanding the quantum workforce and educating the students who will one day become leaders of this industry \cite{fox_preparing_2020,hughes_assessing_2022,aiello_achieving_2021,kaur_defining_2022}. The varied roles in the quantum industry require different levels of education, and one of the necessary pathways into quantum industry is for people with terminal bachelor's degrees.  ~\cite{asfaw_building_2022, perron_quantum_2021}.

However, there are a number of challenges to designing curricula to prepare bachelor's level students to participate in the quantum industry. Because the quantum industry is new, educating a quantum workforce requires the development of dynamic new courses that center on ``complex subject matter'' and ``ever changing instrumentation'' ~\cite{aiello_achieving_2021}. Furthermore, these courses must be able to serve students from a variety of academic backgrounds and must give them the skills to work in many areas of a broad industry ~\cite{aiello_achieving_2021,asfaw_building_2022}.

In response to the need for growth in the area of QIST, the National Science Foundation (NSF) has deployed its Quantum Leap Challenge Institute (QLCI) program, which consists of five institutes across the U.S. seeking to ``advance the frontiers of quantum information science and engineering''~\cite{noauthor_home_nodate, communications_home_nodate, noauthor_nsf_nodate, noauthor_nsf_nodate-1, noauthor_institute_nodate}. One of these institutes, centered at the University of Colorado (CU) Boulder, is called Quantum Systems through Entangled Science and Engineering (Q-SEnSE). In line with the goals of the NQI, one of the desired outcomes of the Q-SEnSE institute is to ``train a quantum savvy workforce''~\cite{noauthor_home_nodate}. There are many ways that Q-SEnSE seeks to accomplish this goal, and one of these methods is a new year-long course at CU Boulder entitled Quantum Forge (Q-Forge), which was first offered in the Fall of 2022 ~\cite{noauthor_home_nodate}.

Capstone courses (or similar types of practical training) have been suggested as an important part of educating students to enter the quantum workforce, especially for students without a graduate degree \cite{fox_preparing_2020, aiello_achieving_2021, asfaw_building_2022}. Q-Forge is an upper-division capstone-style course that allows students to partner with a company in the quantum industry to work on an authentic project for their partner company. It is described fully in Sec. ~\ref{Course Context}.
%
%

Because Q-Forge is a novel type of course in preparing students for participation in the quantum industry within the context of physics, it is important to investigate if it is meeting its goal of introducing students to the quantum workforce and helping them develop the ability to be successful within it. If so, this course could be a model from which to develop similar courses at other institutions, thereby providing additional opportunities for students to obtain the kinds of experiences often historically available only through individual research positions or internships. In order to ensure that courses such as Q-Forge are meeting the needs of all students, research on the impacts of courses, especially on students' self-efficacy, must be done ~\cite{aiello_achieving_2021, asfaw_building_2022}. Thus, our research questions are as follows: 
\begin{itemize}
    \item \textbf{RQ 1}: What do Q-Forge students think that the quantum industry is?
    \item \textbf{RQ 2}: To what extent do Q-Forge students think they can be successful in the quantum industry?
    \item \textbf{RQ 3}: Do Q-Forge students want to participate in the quantum industry? 
\end{itemize}
For each of these research questions, we focus on how student responses to these were shaped by the course throughout the year. We explicitly focus on student perspectives on these questions in order to center the voices of those impacted by the course rather than the course designers or instructors.

In this paper, we begin by presenting some background information (Sec.~\ref{sec:Background}) on the quantum industry landscape, previous efforts to educate a quantum workforce, and the educational benefits to students who have participated in other types of capstone courses. We then provide details on the implementation of Q-Forge in Sec.~\ref{Course Context}, including the class activities and the students' industry project. Section~\ref{Methods} describes the process we conducted to collect and analyze several types of data from the first cohort of Q-Forge students. Our results section (Sec.~\ref{Results}) is split into three subsections, each designed to answer one of our three research questions. Finally, we review the conclusions that can be drawn from this work, provide recommendations to future course instructors, and detail future work to be done on this topic in Section ~\ref{Conclusions}. 

\section{\label{sec:Background}Background}

In this section, we  begin by presenting an overview of the quantum industry so that readers have background on the industry for which students are being prepared to participate. We then move into a discussion about previous methods of educating a quantum workforce, specifically at the undergraduate level. Finally, we briefly describe some benefits students receive from participating in engineering capstone courses, as Q-Forge was modelled after these courses.

\subsection{\label{Landscape} Quantum Industry Landscape}
In order to understand how to better educate a quantum workforce, we must first understand what the quantum industry is, how it functions, and what it is looking for in new employees. Several previous studies, summarized below, have focused on these aspects of the quantum industry within the U.S. and Europe ~\cite{fox_preparing_2020,hughes_assessing_2022,aiello_achieving_2021,kaur_defining_2022, hasanovic_quantum_2022, greinert_future_2023}. 

The quantum industry engages in several distinct activities that all contribute to the ultimate goal of advancing quantum technologies and creating successful quantum products. There are various ways to classify the activities that quantum companies engage in. One classification, as described by Fox et al., divides the endeavors of quantum companies into five categories: Quantum sensors, quantum networking and communication, quantum computing hardware, quantum algorithms and applications, and facilitating (i.e., enabling) technologies \cite{fox_preparing_2020}. 
As educators aim to build courses to prepare students to join all categories of companies in the quantum industry, knowledge about the job prospects for these students is crucial. Across a wide variety of companies, there are various types of technical roles commonly reported within the quantum industry. These include engineers, experimental scientists, theorists, technicians, and application researchers ~\cite{fox_preparing_2020}. Importantly, the job title `engineer' is extremely common within the quantum industry, even for students graduating with physics degrees ~\cite{fox_preparing_2020, hughes_assessing_2022}. Furthermore, the types of jobs that quantum companies are recruiting for in the next two years were approximately the same as those in the next three to five years, as reported by companies in 2022 ~\cite{hughes_assessing_2022}. This indicates that job opportunities are likely to stay the same in the short term within the quantum industry.  

There are various technical skills highlighted by the industry as important for future employees. Due to the variance of companies and job roles within the quantum industry, it is highly likely that different employees will require different skills when they first start and surely as they progress through their careers. Nevertheless, there are some commonalities across the industry. First, reliance on coding or computer programming skills is extremely common among the industry, as is reliance on statistical skills for data analysis ~\cite{fox_preparing_2020}. Other skills mentioned by members of the quantum industry as important are troubleshooting; identifying sources of noise; modeling; and understanding quantum concepts such as error correction, (de)coherence, open system dynamics, qubit hardware, Hamiltonians, linear algebra, and quantum circuit design ~\cite{aiello_achieving_2021, fox_preparing_2020}. Perhaps unexpectedly, ``quantum specific skills,'' or those skills that are unique to the quantum industry and require quantum knowledge, are not necessary for all job classifications, indicating that individuals who are not specifically trained in quantum physics may be able to fill essential roles within the quantum industry. However, many of the jobs that are marketed primarily to physicists within the quantum industry certainly do require quantum specific skills, and thus, courses for physicists, such as Q-Forge, may need to emphasize some of these skills to effectively train students for physicist roles~\cite{hughes_assessing_2022}.

Many of the skills that are highly valued within the quantum industry are experimental skills \cite{fox_preparing_2020, asfaw_building_2022}. In fact, courses like Q-Forge may be of particular value for students entering the workforce with a Bachelor's degree, as students with experience in a capstone project or internship are valued for their knowledge of how to work in a quantum lab. The skills gained in a hands-on capstone course like Q-Forge can then supplement experimental skills gained from more traditional lab courses~\cite{fox_preparing_2020}. Capstone courses and other forms of practical training may be of particular importance for students entering a constantly-evolving field like QIST because they provide students an opportunity to understand what their work life within the industry might be like, as opposed to a laboratory course that may not be able to change with sufficient frequency to capture the dynamic nature of the industry\cite{aiello_achieving_2021}.

Another important aspect to preparing a quantum workforce is understanding what types of degrees and credentials companies require of their employees. The answer to this question is not fully known, although several groups have taken, and are continuing to take, steps towards answering it. Notably, Ph.D.s are the most commonly required degrees for employees entering the quantum industry, followed by Bachelor's degrees and, finally, Master's degrees generally, although it does vary by job type~\cite{fox_preparing_2020, kaur_defining_2022,hughes_assessing_2022, noauthor_nsf_nodate-2}. Furthermore, many jobs that require only a Bachelor's degree also require at least three years of experience, where the required experience decreases for more advanced degrees~\cite{kaur_defining_2022}. This indicates that few individuals are currently being hired into quantum companies directly after graduating with a Bachelor's degree, which is important to keep in mind when preparing undergraduate students to participate in the industry. Finally, the jobs that are available for individuals with only Bachelor's degrees often require a degree in computer science or engineering rather than physics ~\cite{hughes_assessing_2022}.

\subsection{\label{Education} Education for a Quantum Workforce}
While educators and institutions across the U.S. are aware of the need for developing a quantum workforce, there are a myriad of challenges to implementing education for a quantum workforce within Bachelor's and Master's programs. The subject matter to be taught is complex and ever-changing, requiring instructors and graduate student teaching assistants to be up-to-date on the latest technologies. Furthermore, the quantum industry is interested in hiring individuals from a variety of educational backgrounds, particularly from distinct disciplines or majors, indicating that courses geared towards individuals interested in the quantum industry must take into account these varied backgrounds ~\cite{aiello_achieving_2021}. Interdisciplinary courses are also common, meaning that many individuals teaching QIST courses are not necessarily subject-matter experts ~\cite{meyer_todays_2022}.

A number of institutions, however, have taken on this challenge of preparing students for work in the quantum industry through various approaches ~\cite{aiello_achieving_2021, perron_quantum_2021}. These approaches include the development of modules in an existing course, a new course, a concentration or track within a major, a minor or certificate program, or even a standalone QIST major or Master's program such as those described in ~\cite{perron_quantum_2021,aiello_achieving_2021, asfaw_building_2022}, among many others. Many of these efforts focus on Master's degree or Ph.D. students, but some are geared towards Bachelor's degree students as well ~\cite{aiello_achieving_2021,asfaw_building_2022,dzurak_development_2022, cervantes_overview_2021}. There are many examples of recently-developed QIST concentrations, certificates, and minors \cite{perron_quantum_2021, asfaw_building_2022,aiello_achieving_2021}. These require students to take several quantum-based courses in addition to their other coursework. The additional courses range from a course on the foundations of quantum computing to standard quantum mechanics courses to courses about specialized kinds of quantum hardware ~\cite{perron_quantum_2021, aiello_achieving_2021, asfaw_building_2022}.

Institutions are also working hard to develop stand-alone QIST courses. For instance, out of 305 surveyed institutions in the U.S., 74 (or 24\%) have at least one QIST course in their course catalogues ~\cite{cervantes_overview_2021}. 123 separate courses were being taught at these 74 institutions, with 46 (37\%) being explicitly directed towards undergraduates and 15 (12\%) listed as both graduate and undergraduate level courses~\cite{cervantes_overview_2021}. Importantly, these courses were offered through a variety of different departments including physics, electrical/computer engineering, computer science, and mathematics ~\cite{cervantes_overview_2021, meyer_introductory_2024}. Most of these courses are offered at doctoral institutions, but some are also offered at Master's and baccalaureate institutions ~\cite{cervantes_overview_2021, meyer_investigating_2024, plunkett_survey_2020}. This suggests that there is still work to be done in ensuring that quantum focused courses are equally accessible to students from all backgrounds, especially low-income and rural backgrounds, an explicit goal of the NQI ~\cite{cervantes_overview_2021, subcommittee_on_quantum_information_science_committee_on_science_of_the_national_science__technology_council_quantum_2022, noauthor_enabling_nodate, meyer_investigating_2024, meyer_disparities_2024}. 

There are still ongoing conversations about the important elements of quantum education. QIST instructors continue to attend workshops about quantum education and discuss how to best implement QIST education at post-secondary institutions ~\cite{asfaw_building_2022, perron_quantum_2021}. Many suggestions for how to successfully implement small modules, courses, and degree programs have come out of these discussions ~\cite{aiello_achieving_2021, asfaw_building_2022, perron_quantum_2021}. 
These suggestions around how best to implement quantum industry focused courses and programs for undergraduates are varied, ranging from ``develop strong relationships with industry'' to ``introduce students to the field as early as possible'' ~\cite{aiello_achieving_2021, perron_quantum_2021}. Some suggestions at the institutional level are to gather subject-matter experts to teach these courses, provide research experience to students, and establish mechanisms to ensure institutional support ~\cite{aiello_achieving_2021}. Meanwhile, at the course level, literature suggests that it is important to educate students about the breadth of the quantum industry. Many courses currently focus on quantum information and quantum algorithms, but do not delve into quantum hardware or other quantum technologies, despite the importance of this knowledge to the quantum industry ~\cite{asfaw_building_2022}. We must also focus on how to encourage diversity, equity, and inclusion within the field from the beginning of any new program or effort to educate a quantum workforce ~\cite{asfaw_building_2022}. One proposed way to attract and retain a diverse set of students is by improving students' QIST self-efficacy and ensuring we study the impact of new courses and programs on students' self-efficacy\cite{asfaw_building_2022, aiello_achieving_2021}.

\subsection{\label{Engineering Capstones}Engineering Capstone Courses}

Although not typical in physics programs, senior design capstone projects have become increasingly common within engineering programs, especially with the introduction of ABET requirements~\cite{dutson_review_1997, noauthor_home_nodate-1}. They have a goal of introducing students directly to the practical side of engineering as part of their curricular requirements, as well as to prepare them to enter industry after graduation ~\cite{dutson_review_1997, todd_survey_1995}.  Although capstone courses vary significantly across engineering sub-fields, they usually consist of some sort of project that involves design, and students often work in teams to accomplish the goals of their project ~\cite{dutson_review_1997,bauer_review_2012}.

Students have historically found that capstone projects offer great educational benefits ~\cite{todd_survey_1995, howe_2005_2006, joo_project-based_2021, shurin_role_2021, goldberg_benefits_2014, jorgensen_industry_2001, howe_preliminary_2018, christensen_capstone_2003}, especially when the course involves an authentic industry project ~\cite{goldberg_benefits_2014, jorgensen_industry_2001}. Students who participated in authentic industry projects reported that they felt a sense of purpose from participating in the project and that they felt like they were making a difference by solving a real problem ~\cite{goldberg_benefits_2014}. Students also noted that they feel capstone projects have at least some value in preparing them with the skills required to enter industry after graduation~\cite{howe_preliminary_2018, christensen_capstone_2003}.

\section{\label{Course Context} Course Context}
In an effort to meet the community's need for the development of a quantum workforce, Q-SEnSE, along with JILA and the physics department, supported the development of the Q-Forge class at CU Boulder. This course takes place over two semesters so that students have the opportunity to fully engage with a longer-term industry project. The goals of this course are for students to be engaged in an authentic quantum industry project, to build the skills necessary to be a part of the quantum workforce, to become motivated to pursue such a career, and to feel that they are capable of pursuing one. These goals are not explicitly addressed anywhere else in the physics curriculum at CU Boulder, placing Q-Forge in a unique position to assist interested students along their path towards a career in quantum industry. 

CU Boulder began offering the Q-Forge class in the Fall of 2022. Designed to be similar to a traditional engineering senior capstone course, Q-Forge partners teams of students with companies in the quantum industry so that students can work on an authentic project with this industry partner. The industry partner supplies the project, modest funding, and support through interactions with their employees throughout the year. These projects seek to expose students to what a career in quantum industry might look like, help students build relevant skills in an authentic way, and help students build networks with individuals who are already a part of the quantum industry.

As the 2022-2023 academic year was the first year this course was offered, we collected data both for research purposes and to provide feedback to the course instructor. This study focuses on the student experiences from this first year that the course was offered. Small changes have been made to the course for the 2023-2024 academic year, which will be detailed in future work.

In the Fall 2022 semester, eight students were enrolled in Q-Forge, five of whom continued with the course in the Spring 2023 semester. The students who did not continue on to the second semester of Q-Forge graduated in December of 2022. In anticipation of the smaller class size the second semester, all students in the class worked on a single industry project.

The students' industry project involved the optimization of heat exchangers for a new type of dilution refrigerator. Dilution refrigerators are a necessary hardware for some types of quantum computers, meaning that this project falls into the ``enabling technologies'' category of activities within the quantum industry. During the Fall semester, the students developed simulations and Computer-aided Design (CAD) models for their proposed new heat exchangers. At the end of the first semester, the specifics of the project goal were changed dramatically after realizing that the degree of uncertainty involved in the project created mismatched expectations between the students and the company. Once the students received their new project goal, which still focused on optimizing heat exchangers, but included a more specific endpoint, they were able to make progress on it during the Spring semester. During that time, they machined and tested a prototype for their industry partner.

Throughout the academic year, students in the Q-Forge class met twice a week for an hour each time with the course instructor. During this time, students listened to lectures, participated in skill development modules, had visits with local quantum companies, and had time to ask questions of the instructor. These meetings became more informal as the year progressed.
Students also had designated time for approximately three hours after one of these lecture sessions each week to work together on their project.

The first week of class in the Fall 2022 semester was dedicated towards familiarizing students with the course, the syllabus, and the lab space, as well as allowing them to have an introductory meeting with their industry sponsor. In the next several weeks, the instructor gave lectures on important skills for working in the quantum industry including computer programming, measurement uncertainty and linear regression. Students also engaged in ``Essential Skills Modules'' geared towards project management skills that would be useful for working in the quantum industry. Each of these four essential skills modules focused on a different relevant skill, such as group roles; communication; appropriately allocating resources for a long scale project; and project inputs, outputs, outcomes, and impacts. Alongside these Essential Skills Modules, the students received a training on bystander intervention with the goal of equipping students to proactively intervene if group dynamics were causing harm to students.
During the second half of the first semester, students visited, and were visited by, local companies in the quantum industry, so that they could better understand what types of companies and jobs exist in the space. Whenever students were not occupied by these skill building activities or visits with local companies, they were working on their industry project together, either with or without the instructor and TA. This project activity then became their only class requirement during the second semester, as the demands of the project grew. Throughout the year, students were able to work on their project in and outside of formal class time and had access to their lab space at any time to facilitate this level of autonomy. The full schedule for the first semester can be found in Table ~\ref{tab:courseschedule}.

\begin{table*}

    \centering
    \begin{tabular}{P{0.1\linewidth}P{0.4\linewidth}P{0.5\linewidth}} \\ \hline \hline
       Week & Tuesday & Thursday \\ \hline
       Week 1 &  Course expectations \& syllabus & Meeting with industry sponsor, introduction to Q-Forge lab space \\
       Week 2 & Discussion of project & Lecture on measurement uncertainty, lab about measurement uncertainty  \\
       Week 3 & Lecture about fitting functions, especially linear regression & Discussion about working with on-campus shops, intellectual property and nondisclosure agreement discussion, activity about repeatable measurements \\
       Week 4 & Group roles presentation & Meeting with industry sponsor, project planning \\
       Week 5 & Project inputs, outputs, outcomes, and impacts presentation & Clear and Efficient Professional Communication presentation, project planning/work \\
       Week 6 & Bystander intervention training Pt. 1 & Bystander intervention training Pt. 2, trip to industry partner site\\
       Week 7 & Dividing projects into specific tasks, allocating resources presentation Pt. 1 & Career services workshop, project work \\
       Week 8 & Allocating resources presentation Pt. 2 & Project work \\
       Week 9 & Tour of maker space on CU campus & Project work\\
       Week 10 & Project work & Trip to local quantum company, project work \\
       Week 11 & Visit from local quantum company & Industry sponsor check-in, project work \\
       Week 12 & Visit from local quantum company & Industry sponsor check-in, project work \\
       Week 13 & Visit from local quantum company & Project work \\
       Week 14 & Project work & Industry sponsor check-in, project work \\
       Week 15 & Project work & Photovoice focus group (optional), project work \\
        \hline \hline
    \end{tabular}
    \caption{The schedule of the first semester of Q-Forge in the 2022-2023 academic year}
    \label{tab:courseschedule}
\end{table*}

Students also engaged with two types of metacognitive reflection as a part of this course: reflection questions and photovoice. These are discussed further in Sec. ~\ref{Methods}.

\section{\label{Methods} Methods}

\subsection{Data Sources}

The participants in this study include the five students who completed the full year of Q-Forge in the 2022-2023 academic year. We chose to exclude those students who did not complete the entire year so as to see the effects of the full course on the students. The five students that completed the full year course are Reese, Jasper, Stella, Nina, and Owen (all pseudonyms). Reese, Jasper, and Owen identified themselves to us as white men, and Stella and Nina identified themselves to us as white women.

Alongside the more traditional coursework that students completed throughout Q-Forge, we also integrated two forms of metacognitive reflection into the course. This was done both to help with our research and to help the students learn how to reflect on their experiences working on such an open-ended, authentic project. While the researchers were not the instructors of Q-Forge, the course instructor allowed us to incorporate these assignments into the course. Each of these assignments was graded strictly for completion, and the course instructor did not have access to specific student responses to these assignments. We monitored the responses weekly and provided feedback to the instructor when concerns or suggestions arose. Each week, students were required to respond to two or three reflection questions and one photovoice prompt. 

Photovoice is a method of participatory action research in which participants photograph something relevant to their lives and reflect on it through captioning and discussion in order to affect change in their community ~\cite{wang_photovoice_1997, oliver_implementation_2024}. In implementing photovoice, we provided students with a prompt each week, and they took a photo in response to that prompt. They then captioned their photo and submitted it to the researchers via a Qualtrics survey. The photos were then discussed during focus groups where students had the opportunity to talk with one another about the meaning of their photos and how they related to the course. We held one focus group at the end of the Fall 2022 semester and two focus groups throughout the Spring 2023 semester, at the request of the students who valued the focus groups as a time to come together and share their thoughts and opinions ~\cite{oliver_prettiest_2023}.  Our photovoice data consists of the audio and video recordings of the focus groups, as well as the students' photos and  captions.  
A complete list of the photovoice prompts related to the quantum industry and the focus group protocol we followed can be found in the Supplemental Material, and a full description of how we implemented photovoice is in Ref.~\cite{oliver_implementation_2024}.

The weekly reflection questions were designed to allow the students to reflect on their experience in the course over the past week. The responses to these questions were collected on Qualtrics, and varied in length from one sentence to an entire paragraph. Students were required to complete these reflection questions each week. The full list of reflection questions we asked that were related to the quantum industry can be found in the Supplemental Material.

The students also had the opportunity to participate in optional interviews with one of the authors (KAO) outside of class three times throughout the year, at the beginning and  end of the fall semester and at the end of the spring semester. In these interviews, we followed a semi-structured interview protocol. The second and third interview protocols were slightly different than the first, as during the first interview students had not yet experienced the Q-Forge course. The parts of the interview protocols used for the work presented here can be found in Supplemental Information. Questions about the students' demographics were asked the first time we interviewed each student, and students were given the chance to update any of their demographic information in subsequent interviews. Students were compensated for participating in interviews with a \$25 Amazon gift card, and could participate either in person or on Zoom. One student, Jasper, participated in the first interview; three students, Jasper, Nina, and Owen, participated in the second interview; and all five students participated in the final interview. It is important to note that we only have interview data spanning the entire year for Jasper. 

These three sources of data described above are used in this work: interviews, reflection questions, and photovoice. All three data sources asked students questions about their career plans, their experiences with and expectations about teamwork, and their familiarity with, and interest in, the quantum industry.

\subsection{Analysis}

All of the data from interviews, reflection questions, and photovoice photos and captions were analyzed using emergent thematic coding ~\cite{merriam_qualitative_2015}. First, these data were sorted by question based on which research question they could answer, and then KAO coded the sections of these data that answered each research question. This codebook was then edited by KAO and VB, where they collapsed codes that were redundant and added new categories for data that were distinct. Examples of the coding were then examined by all authors and discussed until consensus was reached. Inter-rater reliability (IRR) was conducted by authors KAO and VB using a selection of excerpts. We calculated Cohen's kappa for each code and then averaged the values to achieve an average Cohen's kappa of 0.81, which indicates strong agreement~\cite{mchugh_interrater_2012}. However, the small number of participants in this research limits the significance of any claims of the frequency with which codes appear. Our final codebooks will be presented in Sec.~\ref{Results}. 

In order to analyze the data from the photovoice focus groups, KAO and VB watched the video recordings of the focus groups and, together, created content logs to document when topics of conversation occurred ~\cite{jordan_interaction_1995}. This content log was then used to reference specific moments in the focus group that were relevant to the themes discussed in this paper.  

\subsection{Limitations}

There are several important limitations to this study. First, only five  students participated in the Q-Forge course for the entire 2022-2023 academic year and were therefore eligible to participate in this study. Because of this, the results we obtain may be idiosyncratic to these individual students, although they can give some examples of ways that students may think about the quantum industry after engaging in a course like Q-Forge.

Furthermore, we were able to obtain a first interview from only one student, limiting our abilities to fully identify what views students came into the course with and whether or not they changed throughout the year. We do, however, have reflection question and photovoice data from all students throughout the year, so we are able to identify some student views on the quantum industry from the beginning of the course. 

We also studied a specific course at a specific university in this study. There are many different possible ways to educate students to become a part of the quantum workforce, and Q-Forge is one example of a type of course that can potentially achieve these goals. Students engaging in a different type of course, or belonging to a different college or university, might have vastly different beliefs about the quantum industry than these participants did. 

Finally, in order to preserve participant anonymity, as well as the proprietary information of their partner company, there were certain photovoice photos and captions that we cannot share here. While we made every effort to make photos fully anonymous, some student photovoice contributions contained sensitive details about individuals or the specifics of their project, and therefore they cannot be shared widely. This limited our ability to share only a few photos and captions, none of which would have significantly changed our conclusions.

\section{\label{Results} Results}

Due to the fact that the type of responses students gave to answer ~\textbf{RQ 3} were different in nature than their responses to ~\textbf{RQ 1} and ~\textbf{RQ 2}, we present our answers to the research questions in different ways. For ~\textbf{RQ 1} and ~\textbf{RQ 2}, we examine the most common, surprising, or interesting themes for each research question in detail and discuss what these themes tell us about the students' perception of the quantum industry. Then, to answer ~\textbf{RQ 3}, we present a narrative about each student's level of desire to enter the quantum industry. While the data relevant to ~\textbf{RQ3} were analyzed in the same way as ~\textbf{RQ1} and \textbf{RQ2}, the resulting themes are presented as narratives because the question of whether a student wants to enter the quantum industry is a highly personal one that depends greatly on each student's individual feelings about the industry and personal situation. Therefore, it makes sense to tell each student's story individually to provide examples of how students might feel about entering the quantum industry.

\subsection{\label{RQ1}What do students think that the quantum industry is?}

We begin by presenting our results to ~\textbf{RQ 1}: What do Q-Forge students think that the quantum industry is? In order to answer this question, first, we discuss what the students knew about the quantum industry at the beginning of the course, and then move into a description of several characteristics of the quantum industry as described by the students. Next, we detail what feelings students had about the quantum industry, and we complete our answer to this research question by describing where students learned what they knew at the end of the course about the quantum industry. 

\subsubsection{Students' knowledge of the quantum industry at the beginning of the course}

Student knowledge of the quantum industry at the beginning of the course is difficult to know in detail, as we were able to conduct a pre-interview with only one student. Nevertheless, we do have access to students' reflection questions from the beginning of the year, and we asked them to reflect on this question during subsequent interviews, so we are able to provide some insight into their views on the industry before the course began. 

For the most part, these students did not know very much about the quantum industry before this course began. In the pre-course interview, when asked what he knew about the quantum industry, Jasper said,

\begin{quote}
``Essentially nothing. I've heard a bit about just, like, some of the stigmas about quantum and the hype, and the way that there's lots of hype chasers and people pushing for computing. And in some sense this is a little bit part of that train, I feel like, because the idea that there would be commercially available refrigerators at this quality and scale is definitely banking on the importance of quantum computing in the future, which I'm not totally sold on yet, because I don't fully understand the logistics of it. Yeah, I know a little bit about some friends that work on quantum type stuff, but really nothing.''
\end{quote}

Nina added to this theme in the mid-course interview when asked about what she understands the quantum industry to be in comparison to the beginning of the year:
\begin{quote}
 ``I had, like, no idea what quantum industry kind of entailed. I knew that there was such a thing called a quantum computer, that I didn't understand a ton about. I knew it had qubits, but other than that, didn’t know, like, the different ways to construct one and different, like, I don't know, that that was part of the quantum industry.''   
\end{quote}

Even when students said that they felt knowledgeable about the quantum industry before the course began, their following statements did not always support this idea. For instance, when asked in the mid-course interview about how his knowledge of the quantum industry compared to what he knew before the course, Owen said that he had a ``pretty good idea'' about what the quantum industry entailed, but then went on to say,

\begin{quote}
    ``I didn't know just quite how different each of the approaches really were. The specific of each of the approaches. I think that's what's really different. I kind of knew that everyone had their own way of looking at quantum, but I didn't know just how unique they all were and the different, kind of, challenges each approach involves.''
\end{quote}

These quotes illustrate that many upper-division physics students, even those who chose to take a course centered on the quantum industry, may be unfamiliar with what types of activities are occurring within it. For this reason, it is likely important that courses similar to Q-Forge seek to educate students about the industry as a whole alongside preparing them to take part in it, so that students entering the quantum workforce are knowledgeable about what it does, what companies exist, and what jobs are open and accessible to them. Furthermore, any given capstone-style project that students work on in a course like Q-Forge will be able to give students exposure to only one aspect of a very diverse industry, highlighting the need for further education about the variety of work done in quantum spaces.

\subsubsection{Aspects of the quantum industry}

Once the students had taken part in the Q-Forge course and became more familiar with the quantum industry, they shared with us some of the aspects of quantum industry that were particularly salient to them. The first of these ideas is that there are many ways to engage in the quantum industry, and that the quantum industry is not just quantum computing. This idea was brought up by four out of the five students at some point throughout the course, with the first mention of this theme occurring at the end of the first semester of the course. For instance, when asked about what she understood the quantum industry to be, Nina said,

\begin{quote}
    ``it's not just quantum computing, like, there’s signaling, and, like hardware that goes, you know, that's considered quantum industry and quantum cryptography and quantum this and that and just a bunch of things that I didn't really know were things that are actually being worked on.''
\end{quote}

This awareness of the breadth of the industry may have come in part because the students' partner company is aiming to become a leader in dilution refrigerators, which are an important part of the support infrastructure for other quantum systems. Through their project, these students were therefore exposed to this crucial piece of quantum industry, which made them aware of the variety of work that quantum companies can do. When asked at the end of the year about whether he had more knowledge of what jobs exist in the quantum industry, Owen specifically described the role that the project played in changing his viewpoint:
\begin{quote}
    ``at the start of the semester, when I thought about quantum computing, I thought about Hadamard gates and executing code and running Shor's algorithm, and, you know what this project has done, it's opened my eyes to the fact that there's also a lot of other preliminary stuff that needs to happen before we can quantum compute and run our Hadamard gates. And that's where all the jobs are. There's like a lots, you know, it's not just- there's a lot more to it.
\end{quote}

If one of the goals of a course to educate the quantum workforce is to expose students to the variety of types of activities within the quantum industry, it may be useful to have them work on a project outside of quantum computing so that they can experience firsthand the breadth of the industry. 

Another characteristic of the quantum industry that stood out to the Q-Forge students is that it lacks diversity. While creating a diverse and inclusive workforce is one of the goals of the NQI, it is important to note that white men are vastly over represented in the disciplines of physics and engineering more broadly ~\cite{noauthor_minority_nodate,roy_engineering_2019}. Therefore, it would make sense for the quantum subspace to bring with it the lack of diversity of physics, engineering, and computer science as a whole. While the quantum industry is still quite young, this lack of diversity is already showing up in a way that is obvious to students who will be the next generation of the quantum workforce. 

Jasper highlights the issue as he sees it in his interview at the end of the year by saying in response to a question about what he understands the quantum industry to be,

\begin{quote}
    ``I think the work [in the quantum industry] is all really cool, and it's very interesting, which I think is really where the hype comes from, but I think as a result of that, there's a lot of things that have been left behind by the industry, like one thing in particular that I've been kind of, I guess disappointed by is that the places that we visited have been either predominantly or entirely composed of white men, which, I mean, I know that's kind of just physics as a field in some sense. But I guess, I think it's also, kind of, in some ways, a consequence of that hype and the excitement where it's just: we have to make progress, because in their eyes, in the eyes of these people who are putting money in this thing and putting time and investing their lives in this- this industry, quantum is the future.
\end{quote}

Jasper explained that a potential reason behind the lack of diversity in the field is that those who are in charge find making progress more important than making the field equitable. The validity of this assumption is uncertain, but it is important to note that students may gather this impression from the lack of representation they see around them. 

While these were two interesting characteristics of the quantum industry that were brought up by the students, they are far from the only ones. Table ~\ref{tab:aspectsofindustry} describes the other aspects of the industry that the students discussed in our data sources. 

\begin{table*}
\caption{\label{tab:aspectsofindustry} Characteristics of the quantum industry as brought up by the Q-Forge students. We do not provide any information about the frequency with which these codes occur, as the presence of these characteristics simply provide us with examples of what students might notice about the quantum industry. Characteristics that are discussed in more detail in the body of the paper in this section are bolded.}
    \begin{tabular}{P{0.3\linewidth} P{0.7\linewidth}}
        \hline\hline
        Aspect of QI & Description \\
        \hline
         \textbf{Breadth of quantum industry}  & Student feels like there are lots of different ways to engage in quantum industry. \\
         Design & Student feels like creating the design for a project is an important part of work in the quantum industry. \\
         Hype & Student feels like there is a lot of hype surrounding the quantum industry. \\
          Lab work & Student feels like lab work is an important part of work in the quantum industry.\\
         \textbf{Lacks diversity} & Student feels like the quantum industry lacks racial/gender diversity.\\  
         Less complicated & Student feels like the quantum industry is less complicated than the student originally thought.\\
         Many unknowns & Student feels like there are many unknowns about the future of quantum industry and how to achieve the industry's goals. \\
         Mostly startups & Student feels like the quantum industry consists of mostly startup companies.\\
         Optics & Student feels like the quantum industry often involves work on optics. \\
         Requires a Ph.D. & Student feels like working in the quantum industry requires a Ph.D. \\
        Requires teamwork & Student feels like work in the quantum industry requires teamwork.\\
         Specialized knowledge & Student feels like work in the quantum industry requires specialized knowledge.\\
         Work on Authentic Projects & Student feels like working on authentic projects, such as their industry-sponsored project, is similar to work in the quantum industry.\\
             \hline\hline
    \end{tabular}
\end{table*}

Table~\ref{tab:aspectsofindustry} captures ideas that appeared at least once within our students' data. We do not aim to make any claims about the frequency with which these aspects occur, nor do we expect that a different group of students would have provided the same list. Instead, Table~\ref{tab:aspectsofindustry} shows us examples of what types of aspects of the quantum industry might be salient to students. Further research is required to understand which groups of students notice these particular aspects, and what specifically led to students noticing each.

Table ~\ref{tab:aspectsofindustry} includes the idea from students that a Ph.D. is required to work in quantum industry. While we do not discuss this in detail here, we go into more depth about this idea in Sec.~\ref{RQ2}, where we discuss reasons that students thought they could not be successful in the quantum industry, as well as in Sec.~\ref{RQ3}, where we discuss each student's reasons for wanting or not wanting to work in the quantum industry.

\subsubsection{Sentiments about quantum industry}

Students also provided us with information about what emotions they felt might be involved in the quantum industry. The main feeling that came up for them was excitement, with several students mentioning the fact that the work the industry is doing is exciting or that the people doing it are excited about it. For example, in an interview at the end of the first semester in response to a question about what parts of the industry he would find interesting, Owen said about the industry itself,

\begin{quote}
    ``What I think is super interesting about [the quantum industry] is the problems that quantum computing can solve. Just what I think is so exciting about quantum computing in general. You know protein folding for drug synthesis and cracking RSA. Who knows if it'll be done in our lifetime. But that's what I find interesting about the whole thing.''
\end{quote}

\begin{figure}[htbp!]
    \includegraphics[width=0.4\textwidth]{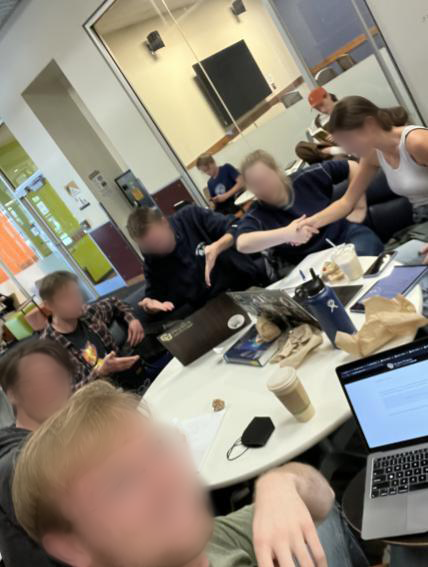}
    \caption{A photo in response to the prompt, ``Take a photo that represents the environment/culture of a company in the quantum industry,'' of the group of students working on a quantum mechanics homework in the library. Jasper's quote reads, ``This photo, a circle of us all working on quantum mechanics homeworks, is a great fit for this prompt. Quantum is a new industry, and is full of excited young people learning and doing new physics.''}
    \label{fig:excitedyoungpeople}
\end{figure}

In response to one of the photovoice prompts during the first semester, ``take a photo that represents the environment/culture of a company in the quantum industry,''  Jasper mentions the excitement of the people working in the quantum industry. Figure ~\ref{fig:excitedyoungpeople} shows Jasper's response to this prompt, and his caption reads, 

\begin{quote}
    ``This photo, a circle of us all working on quantum mechanics homeworks, is a great fit for this prompt. Quantum is a new industry, and is full of excited young people learning and doing new physics.''
\end{quote}

Jasper and Owen both describe feelings of excitement related to the quantum industry, and they both mention the excitement of being involved in something novel. We might expect students who self-selected into a class about the quantum industry to be particularly excited about the industry, but it is beneficial to know that the potential future employees of this industry see at least some degree of excitement within the industry. 

Excitement was not the only emotion mentioned by the students, however. They also mentioned feelings of frustration, isolation, pressure, or that working in the quantum industry might be tedious. Sometimes, students thought that these different emotions could occur simultaneously within them or others. Nina, for example, felt that the quantum industry might be exciting and also frustrating. When asked what she thought working in the quantum industry might be like at the end of the first semester, she said

\begin{quote}
    ``I feel like I still see it as on the edge of science, like, super exciting, but frustrating. I guess most quantum companies are all trying to race ahead of each other right now. It's kind of, like, who's going to be the first to make an actually usable product. And so I feel like I see some competition and some exciting, like, excitement, and then frustration.''
\end{quote}

This goes to show that despite the feelings of excitement students may have surrounding the industry, there may also be some negative feelings about the industry. In order to address these feelings, providing students an opportunity to authentically engage with the industry and decide if the industry is a place for them is important. After all, students learning that they do not want to participate in the quantum industry is in some ways just as valuable as learning that they do want to participate, and it is the role of the individual student to decide if the positives of working in the industry outweigh the negatives.

\subsubsection{Where students learned about the quantum industry}

Finally, we discuss where students learned what they now know about the quantum industry. This may help others decide which aspects of the Q-Forge course to replicate in their own quantum industry focused courses to help students meet their learning goals. 

Stella explained that she learned a lot about other companies in the quantum industry through the interactions she had with these companies during visits to or from the companies. When asked in the post-course interview if she felt like she was more familiar with what companies exist in the quantum industry, she said

\begin{quote}
    ``So, yeah, I feel like I definitely learned a lot of- I found out, I guess, about all the other companies that existed in the realm of quantum industry. And that was really cool. I liked visiting all of the different places, it was nice to get a feel for all of them and see what made each unique and different. And they all had something different they were trying to do in quantum industry and I thought that was pretty cool.''
\end{quote}

Owen also discussed, in the post-course interview, the role of company visits during the first semester in his process of learning about the quantum industry, 

\begin{quote}
    ``The tours that we did last semester I think work great. It made it really concrete that there's a lot of really cool stuff going on. Right- right in our backyards. Yeah, I don't think we've connected as much with local quantum computing companies, you know, we did go on a NIST tour this semester, which was super, super cool. I was a big fan of that. Yeah, I think- I think my perceptions of the industry have stayed about the same, this semester anyway. Last semester's tours were really insightful, though.
\end{quote}

Both of these students felt that they learned a lot from interacting with companies besides their industry partner throughout the first semester. This indicates that industry visits may be an excellent way for a quantum industry focused program to introduce students to the variety of companies within the quantum industry and get them interested in the quantum companies in their local area. Virtual visits could also be used in locations that are not home to local quantum companies. 

While the students felt that they learned a lot about what types of companies exist in the quantum industry, only one student talked about learning more about what particular job roles exist within the industry. Many students responded similarly to Stella, who indicated that she did not know much about particular job roles within the industry. When asked in the post-course interview if she was more familiar with the types of jobs that exist in the quantum industry, Stella said, 

\begin{quote}
    ``I feel like for us, we've seen mostly the mechanical engineering side of things. So I know that there's mechanical engineering, but we don't really know about much else. So I would say, I don't think I've really learned more about the different types of jobs. I just- I know they're there, because we visited other companies. But for our project, we've been working on one thing, so it's hard to be like, I don't have personal experience with, like, those other jobs.
\end{quote}

For Stella, her lack of personal experience with other job roles made it hard for her to imagine what other types of jobs might exist. This may be a challenge to overcome if one of our goals is to educate students about the way they might engage in the quantum industry in the future. Certainly, this goal must be balanced with other course related goals (such as completing an authentic project), but it also may be useful to have companies discuss potential jobs in greater detail if company visits are included in a course or program.  

The one student who did mention feeling more familiar with what jobs exist in the quantum industry, Nina, said in response to a question about what jobs she was familiar with in the quantum industry,

\begin{quote}
    ``...quantum physicists and researchers versus someone in quantum computing or engineering and then, like, there's mechanical engineering roles and electrical engineering roles, and then just, probably, like, analysts and people who are really doing, like, numbers- or analyzing the effectiveness of things.
\end{quote}

Even Nina did not elaborate in depth about the types of jobs she might be interested in looking for in the future, or the specific role of these jobs, and she was the only student to mention feeling like she did know more about jobs after taking Q-Forge. This suggests that it might be beneficial to add more opportunities for students to learn about specific job roles, if this is a goal of a course like Q-Forge. 

Students entered the Q-Forge course with relatively little knowledge about the quantum industry as a whole and then learned throughout the course about several important aspects of the industry. This included learning about the breadth and lack of diversity in the industry and discussing the emotions they recognized within and about that industry. Students also identified where their new knowledge about the quantum industry came from, particularly from visits with companies.

\subsection{\label{RQ2}Do students think they can be successful in the quantum industry?}

~\textbf{RQ 2} investigates whether the students in the Q-Forge course felt that they could be successful in the quantum industry. Throughout the course, there were times when the students felt they could be successful in the quantum industry and other times when they did not. These came about for different reasons, so in this section we consider the various reasons for which students felt they could and count not be successful in the quantum industry.

\subsubsection{Reasons why students thought that they could not be successful in quantum industry}

We begin with the reasons that students felt that they could not be successful in the quantum industry, particularly in terms of obtaining a job directly after graduation. There were several things that students believed might contribute to their lack of success in the quantum industry. We describe two of the most common themes in detail and provide a list of the other reasons at the end of this section. 

The first reason students had for thinking that they could not be successful in the quantum industry is that they noticed the frequency of individuals with Master's degrees or Ph.D.s in the quantum industry and came away with the impression that it is difficult to be successful in the industry without an advanced degree. It is important to note that all five of the students who took Q-Forge mentioned this at one time or another throughout the course. 

When asked about how confident he was in being able to get a job in the quantum industry at the end of the year, Reese said,

\begin{quote}
    ``I haven't, like, dipped my toes in it, but... not very just because I do know that they- they- they want, like, Master's or Ph.D. level, kind of, knowledge and specialized knowledge specifically. Like, I guess in the sense that, like, I took advanced lab [a separate traditional advanced lab course at CU] solely because [individual from industry partner] basically told me that he, kind of, looks for that in a job.
\end{quote}

In this quote, Reese explains that he ``knows'' that quantum companies are looking for the kind of knowledge one acquires after pursuing an advanced degree, and says that he was ``not very'' confident about being able to find a job in the quantum industry with his Bachelor's degree. 

Stella agreed with Reese's perspective on being able to get a job in the industry in her post-course interview, explaining, 

\begin{quote}
    ``Like, I think there's a lot of great things that I have on my resume now because of the project. Like, great resume builder. It's been wonderful. But I'm also not confident about getting a job, especially because a lot of them are looking specifically for either interns, or Ph.D.s and not in between and, like, okay, well, I have a Bachelor's degree, you want to do something with that? And they're like, no, but just- just based on what I've seen.''
\end{quote}

This topic came up in the photovoice focus group in the middle of the second-semester as well, when Jasper acknowledged that quantum companies do sometimes hire individuals with Bachelor's degrees, but that they are hired to do work that is perceived as less valuable or prestigious. Jasper said,

\begin{quote}
    ``It does seem like a lot of the places that we visited and talked to sort of have this pressure of, like, well we do sometimes hire undergraduates, but the only people who are really valuable to our company are the grads.''
\end{quote} 

These three students, alongside Nina and Owen whose perspectives are represented here although they are not quoted directly, saw that the quantum industry is dominated by people with graduate degrees. The literature supports the idea that there are more jobs available for workers with advanced degrees, but there are also some jobs available for workers with Bachelor's degrees ~\cite{kaur_defining_2022, fox_preparing_2020}. More research is required in order to understand in more detail what types of jobs are available to individuals with a recent Bachelor's degree and to develop more knowledge about the accuracy of these students' perspectives that Bachelor's degrees were not of value to quantum companies. Depending on the results of this research, courses introducing students to the quantum industry may need to advertise in a way that accurately portrays career prospects to their students.

There are several ways for courses like Q-Forge to address the perceived lower demand in the quantum industry for students with Bachelor's degrees and no additional job experience. Classes like Q-Forge could further emphasize job preparation such as resume workshops, career fairs, and networking opportunities so that the students participating in the course are well equipped for the job application process. Additionally, it may be helpful to make students aware of the statistics regarding individuals with Bachelor's degrees in the quantum workforce, and to expose them to such individuals within their courses. For example, ensuring that students have the opportunity to speak to someone whose highest degree is a Bachelor's degree during visits to companies may help students develop familiarity with the types of roles that would be directly accessible to them. 

Alongside feeling like they needed a graduate degree, with some students specifying a Master's and other's specifying a PhD, to be successful, students also described feeling like they lacked certain skills required for participation in the quantum industry. Particularly, Jasper said,

\begin{quote}
    ``...since we ended up working on a project that's more focused on the support infrastructure than the actual quantum science itself, it might be that we should try marketing ourselves differently, more as that support role than as physicists, because in a sense, we just don't have that PhD level, like, optics, PhD, like, experience that most, I think, of these companies are really looking for, to do, like, the physics work at the front, sort of edge of that.''
\end{quote}

He felt that participating in a role in a company within enabling quantum technologies meant that he would be unqualified to change his focus if he wanted to engage in work more on the ``actual quantum science'' side of the industry. This is, perhaps, a downside of having students work in a part of the industry that is dedicated to providing enabling technologies for QIST  instead of more directly working on those activities themselves. 

Additional research focusing on the needs of the quantum industry with respect to what skills students need to have when they graduate as opposed to what skills students can learn on the job may be beneficial to understanding whether students need to learn more directly quantum related skills in order to enter the industry directly after graduation ~\cite{noauthor_nsf_nodate-2}. There is also a tradeoff, however, between teaching students skills that are important for the quantum industry and giving them the opportunity to practice these skills through an authentic industry project. The goals of the course, including career preparation alongside job training, must be thoroughly considered in order to determine which activities will best allow the course to meet its goals. 

Other things that contributed to students feeling like they could not be successful included not doing enough job preparation (preparing and reviewing resumes, attending career fairs, etc.) and lacking the necessary connections to individuals within the industry.

\subsubsection{Reasons that students thought they could be successful in the quantum industry}

There were also plenty of reasons why students thought that they could be successful within the quantum industry. While students were concerned that they did not have all of the required skills to participate, students also felt that they did gain important and relevant skills. They also felt that working on an authentic industry project prepared them for success in the quantum industry if they were to get a job. 

When asked if she thought she had the skills to participate in the quantum industry, Stella said,

\begin{quote}
    ``Yeah, I'd say yes. I think- I think I have enough skills that I could get by and I could- I think I have the ability to learn a lot in a short amount of time, so I feel like I have, yeah, I think I- I have the skills for the job. I just don't think I would get one.
\end{quote}

Similarly, Owen felt that he had gained a lot of skills that would be beneficial in the quantum industry. He said,

\begin{quote}
    ``This class has taught me a lot about a bunch of different things, you know. My programming has gotten- has gotten so much better, my programming has improved so much. The machining aspect of things, I've learned so much about working with copper. And even just in- machining in general, which I think could be super valuable to a company. The leadership and interpersonal skills, super valuable. I think I'd be a great addition to somebody's team.''
\end{quote}

Stella's quote indicates that she feels that she would be successful if she were able to get a job due to the skills that she has, and Owen mentions several skills that he learned throughout the course. Certainly, a one-year class like Q-Forge cannot possibly provide students with every skill required for success in the industry, but it is a positive sign that students are leaving the course feeling prepared and confident, at least in some ways. 

Students also felt that working on the authentic industry project prepared them to be successful in the quantum industry. In the post-course interview, Jasper said, 

\begin{quote}
     ``I guess I feel like the best support you can give a student to get a job they want is experience in the relevant area. And so in that sense, I feel like just by doing the project is a great support for understanding, like, helping us push toward knowing what we would need to do to get a job.''
\end{quote}

He felt that working on their project with the industry partner was relevant and that it would provide them with the necessary experience in order to get a job. Jasper also said that this is the ``best support'' one can get as far as job preparation goes, highlighting the importance of having students participate in an authentic industry project rather than simply teaching skills in isolation. 

Stella also discussed this idea in response to a reflection question near the start of the spring semester. When asked what she had done in the past week that helped prepare her for a job in the quantum industry, Stella said,

\begin{quote}
    ``We have run a few experiments and drawn what we feel are meaningful conclusions. While I have already done experiments throughout my college experience, this was one where we came up with the setup and methods of measurement all on our own. Designing our own experiment was very refreshing.''
\end{quote}

While Stella does not directly mention working on the authentic project as a contributing factor to her feeling that she could be successful, she describes her experience with these experiments as different than her other lab experiences in college. 
Furthermore, Stella discusses her team coming up with their own experimental design, a characteristic of an authentic project, and finding that ``refreshing.''

Both of these students' responses highlight the value of students participating in an authentic project with a company in the quantum industry in order to build feelings of confidence within students. This type of course, containing an authentic industry experience, may be of use to students interested in entering all types of industry, not solely quantum. 

There were also other things that led students to believe that they could be successful in the quantum industry. These included: having prior experience in some type of physics related industry, feeling a sense of belonging in the quantum industry, feeling like they were adequately prepared to participate in teamwork in the industry, and feeling like they were able to learn from visits to and from quantum companies. 

\subsection{\label{RQ3}Do students want to participate in the quantum industry?}

Finally, we turn to ~\textbf{RQ 3} and investigate students' individual interest in becoming a part of the quantum industry. In this section, we will address each student's responses to the reflection questions, interviews, and photovoice individually, so as to share the idiosyncratic and deeply personal nature of each of these students' perspectives. We note once again that each student is likely to have their own reasons for wanting or not wanting to enter the quantum industry, and these results are simply examples of what students might feel about joining the industry after participating in an experience like Q-Forge.  

\subsubsection{\label{Reese}Reese}

Reese identified himself to us as a white man. He only participated in the third interview with the researchers, but answered reflection questions and submitted photovoice responses throughout the year. Reese indicated to us that he was interested in participating in the quantum industry, but this interest was not unique to quantum specifically; he was interested in participating in a variety of physics industries. 

\begin{figure}[htbp!]
    \includegraphics[width=0.4\textwidth]{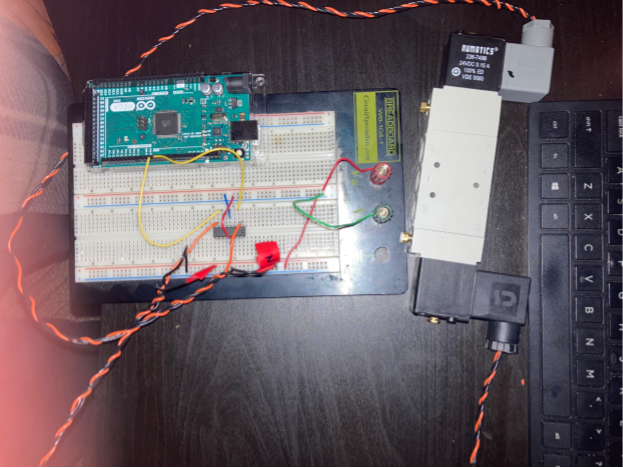}
    \caption{A photo of a breadboard prototype of a circuit. Reese submitted this photo in response to the prompt, ``Take a photo representing your level of interest at the moment about pursuing a career in quantum industry'' with the caption: ``This is my first attempt of building an amplifying circuit to drive out solenoid valve. Definitely still very interested in industry, I love the process of manually prototyping circuits on the bread board which I enjoy!''}
    \label{fig:prototyping}
\end{figure}

Important to Reese's overall perspective on the industry is his love for working with his hands on tangible projects. In response to a photovoice prompt at the end of the second semester that read ``Take a photo representing your level of interest at the moment about pursuing a career in quantum industry,'' Reese submitted the photo shown in Fig. ~\ref{fig:prototyping}. He captioned the photo,

\begin{quote}
    ``This is my first attempt of building an amplifying circuit to drive [our] solenoid valve. Definitely still very interested in industry, I love the process of manually prototyping circuits on the bread board which I enjoy!''
\end{quote}

Through this photo, Reese suggests that he is interested in industry at least in part because he expects to be able to work on tangible, hands-on processes such as prototyping circuits on a breadboard. This ties into the fact that the authentic industry project drove Reese's interest in quantum industry, which will be discussed further below. 

During his interview at the end of the year, Reese also discussed his interest in quantum, as well as other types of industry. He said,

\begin{quote}
    ``It's really good knowing that I could go to quantum industry or aerospace industry or more or less anything with, like, a basic physics degree and- and a little bit of engineering, you know, and that can more or less be anywhere, like, bio, or med, or aerospace, or applied new physics. But it's just- I think it's more of the work environment is where I really want to be picky. Because that's where I think it'll matter the most. Because at the end of the day you're- you may not like your job at some point or another, but the people are what's gonna make it enjoyable to, like, get through the tough times and whatnot.''
\end{quote}

Reese emphasized through this quote that he feels he could be interested in a variety of different physics related industries and that what matters most to him is the culture of the company he ends up working for. Creating environments that are positive will likely attract candidates for whom this is a priority, like Reese. 

Reese also noted that working on the industry project increased his interest in participating in the quantum industry. In his end-of-year interview, Reese once again highlighted the importance to him of getting to work with his hands in a future job, which he did as a part of this industry project. In response to a question asking what parts of the Q-Forge course influenced his interest in quantum industry, Reese said

\begin{quote}
    ``I would say just the amount of machining that I did. I thought it was gonna be a lot more theoretical in that sense of like a lot more computer work, desk jockey kind of stuff, and I wouldn't get out there and do anything kind of deal. My biggest fear is, like, just research and just, like, you know, never actually being integrated or integrating anything or just, like, just always behind that desk and typing away and theorizing but never actually, like, testing and learning and all that sense of, like, physically, like, fucking up and understanding why I did that, you know?''
\end{quote}

The fact that the industry project involved a large amount of hands-on work, specifically involving machining, was a benefit of the project for Reese and helped him become more interested in the quantum industry overall. This is because he learned through his industry project that there are places in the industry where hands-on work is necessary, and that working in the industry would not necessarily be a job where he was behind a desk working on a computer for the majority of the time. Teaching students about the variety of different types of work in the quantum industry may be an important role of courses such as Q-Forge, especially since the industry is, as the students mentioned, quite broad and there are many different types of roles that may be available to these types of graduates.

Additionally, in response to a reflection question at the end of the fall semester asking ``how has working on this project impacted your interest in a career in quantum industry?'' Reese replied,

\begin{quote}
    ``I already wanted to work in the industry, but this experience has given me a better idea of what it takes to make that happen and how I can be a better employee and coworker.''
\end{quote}

While this quote does not directly respond to how Reese's interest in quantum industry increased due to the project he was working on, he stated this in response to a prompt that asked directly about interest. Therefore, we can understand that the project helped Reese gain skills that he felt he needed to be a part of the industry, which reaffirmed his interest in pursuing such a job after graduation. This once again speaks to the power of an authentic industry experience in courses such as Q-Forge that can help students get a better understanding of how to engage with the quantum industry, thereby increasing their desire to do so. 

Finally, Reese responded to another reflection question during the first semester asking about his interest in quantum industry. The question asked, ``after starting to work on the project, how has your interest in quantum industry changed?'' Reese responded,

\begin{quote}
    ``Drastically increased! I am really enjoying the fact that all we are learning now will be integrated into the design requirements of our project. As an engineer, I live for stuff like this.''
\end{quote}

Reese notes that he valued that he was not learning simply for learning's sake, but that he got to apply his knowledge into a real-world project. Allowing students to get experiences applying their knowledge to something important to individuals outside of the course may, at least in part, fuel some students' interest in going into the quantum industry.  

\subsubsection{\label{Jasper}Jasper}

Jasper also identified himself to us as a white man. He was interested from the beginning of the course in attending graduate school. During the year, Jasper was accepted into a prestigious physics Ph.D. program. While Jasper's intentions were never to go into quantum industry, he had some interesting insights on why he was not interested, as well as some aspects of the industry that did pique his interest. Jasper participated in all three interviews, as well as reflection questions and photovoice throughout the year. 

As mentioned above, Jasper entered the course with a plan to go to graduate school and get a Ph.D. in physics. While, typically, Q-Forge would only accept students who are interested in going into the workforce directly after their Bachelor's degree, this first year of the course an exception was made for Jasper. When asked about his interest in quantum industry in the final interview, Jasper said,

\begin{quote}
  ``I guess in some sense I cheated a little bit, because I've always really intended on going to graduate school. And I did make that clear in my application to join the class. And so I- I haven't been trying to hide that necessarily.''
\end{quote}

While it is important to note that Jasper was unique in this course in having little interest in going into the quantum industry directly after finishing his Bachelor's degree, he still had valuable reflections on how the industry functions that encouraged or discouraged his interest. 

One important theme for Jasper that discouraged his interest in quantum industry was the lack of racial and gender diversity that he saw within the industry. During his post-course interview, after discussing the barriers he imagined his classmates who were women might face, Jasper said,

\begin{quote}
    ``...being a woman in this field, or being a person of color in this field, is just more difficult, and that on its own is a qualification. And I think a lot of companies that we talked to give me the impression that they don't realize that and they don't understand the value that diversity brings to a team. And so for me, personally, diversity has been a very big part of my life experience, and learning and growing, I grew up in a sort of immigrant suburb, and I very quickly realized my privilege and understood and contextualized that within the context of my friends who did not have the same privileges. And I would want to continue living in a place where I'm challenged to continue growing and progressing my ideas of how the world really is, and how it should be, how we could get there. And I feel like quantum industry, at least right now, doesn't have that for me.''
\end{quote}

As a student who grew up in a very diverse community, Jasper was interested in continuing to work in a community that reflected this history. He got the impression throughout this course that diversity within the field was less important to its participants than progress in quantum technologies, which would have discouraged him from participating in the industry regardless of his desires to attend graduate school. As mentioned earlier, it is important to note that, despite being relatively new, the quantum industry is already reflecting some of the lack of representation that is present within physics, engineering, and computer science as a whole, and this is noticeable to students. Increasing diversity in the field is one of the goals of the NQI, and also may be deeply important in bringing new talent into the field. 

Jasper noticed that the quantum industry is not necessarily distinct from physics as a whole in terms of its lack of representation. In response to a reflection question at the end of the year asking how the project impacted his desire to enter the quantum industry, he said,

\begin{quote}
    ``I came into the class with a genuine curiosity about the industry, and that curiosity has definitely been significantly helped by all the explorations of the class. That said, I think my interest in the industry hasn't changed much, I don't feel more or less compelled to pursue a career working for one of these companies. I think this is because while many of the things we've found have been positive -- lots of really cool and exciting technology as a focus for a company is really exciting -- many of the things we've seen have also been somewhat sobering. Working for [the industry partner] gave a good taste of work at a startup, and the lack of employee diversity at the companies we've visited have been more than a little sobering. I wish physics, as a field, was more inclusive.''
\end{quote}

While a course like Q-Forge does not have the power to change the representation in the field on its own, it does seem important to recognize the issues that these students see in the field and engage in conversations about them. We gave this feedback to the Q-Forge course instructor who integrated conversations about the state of diversity within the field into the course in the 2023-2024 academic year. Future work will investigate the outcomes of this addition to the course and make preliminary recommendations for how to effectively conduct these conversations among students. 

Despite the lack of diversity in the quantum industry discouraging Jasper from participating in the industry, he also noticed that there were a lot of novel and exciting things going on within the industry that were interesting to him. During the pre-course interview, Jasper elaborated on his feelings of excitement surrounding the industry: 

\begin{quote}
    ``I mean, the newness of it is exciting. It's- it's kind of- I mean, we, we hear all the time about how transistors have revolutionized, like, everything. And, while I don't know if quantum is necessarily the next transistor revolution in the way that lots of people seem to think it is, um, I'm sure people thought the same thing about transistors at the time, too. So it's- it's impossible to say, really, but it could be exciting to be part of something new.''
\end{quote}

Jasper noted that the possibility of creating something new and useful was exciting to him, reminding us that Jasper's lack of desire to go into quantum industry did not equate to a lack of interest in it.

He also compared his interest in the quantum industry to his interest in physics research through a response to a photovoice prompt. Due to the sensitive nature of the photo that Jasper submitted to this prompt, which was of a diagram related to their industry project, we present only the caption here. Jasper said in response to the prompt ``take a photo representing what you find most interesting at the moment about pursuing a career in quantum industry,''

\begin{quote}
    ``I feel that the energy everyone had in this meeting is definitely emblematic of quantum industry as a whole -- it's sort of just a bunch of people trying to turn cool physics into profit, which is really cool in a lot of ways. Sometimes, everyone gets really excited about an idea that seems like it might have a positive effect on the system; it feels like a research breakthrough in a way.''
\end{quote}

Here, Jasper drew a parallel between the breakthroughs that can occur in industry and the breakthroughs that occur in research. In a way, Jasper did not consider these things so distinct from one another. In order to encourage interest in the quantum industry in a new generation of physicists, it may be important to make students aware of these ``breakthroughs'' that can occur within industry, especially within an industry so young. Driving students' excitement about the possibilities within quantum industry may be an important piece of encouraging interest in the industry. 

\subsubsection{\label{Stella}Stella}

Stella identified herself to us as a white woman. She became less interested in entering the quantum industry throughout this course, beginning with a desire to enter the quantum industry and ending, for various reasons, feeling as though it was not what she wanted to do. Stella participated in only an end of the year interview, but she responded to reflection questions and photovoice prompts throughout the course. 

Towards the end of the first semester, we received our first indication both that Stella was originally interested in quantum industry and that that interest was fading due to the change in the industry project that happened as the industry partner noticed that the scope of the original project was too broad for the students. In a reflection question about the impact of the project on the students' desire to enter the quantum industry, Stella wrote,

\begin{quote}
    ``While I still want to pursue a career in industry, the project has put me a little on the fence. The project changed so much within the semester and I felt that the expectations from [the industry partner] (not our instructors) were unreasonable.''
\end{quote}

This indicates to us that Stella was still at least partially interested in entering the quantum industry around the end of the first semester, but that the change associated with the industry project was challenging for Stella in a way that made her less likely to want to enter the industry. While this sentiment was also echoed by others in the first focus group, it is interesting to note that the project change was offputting to Stella, and a project that was more reasonably scoped from the beginning may have been more attractive to her. As future course designers begin to look for appropriate industry projects for courses that are similar to Q-Forge, it may be important to keep in mind an appropriate scope for the project so that students feel successful and accomplished throughout the course rather than overly discouraged.  

\begin{figure}[htbp!]
    \includegraphics[width=0.4\textwidth]{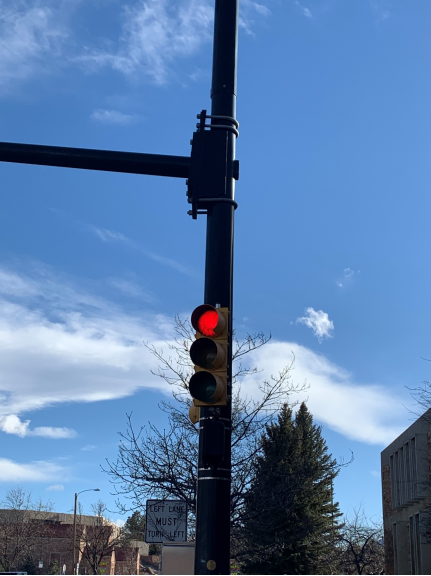}
    \caption{A photo of a red traffic light submitted in response to the photovoice prompt ``Take a photo representing your level of interest at the moment about pursuing a career in quantum industry.'' This photo was captioned: ``As of right now, it's a no-go.''}
    \label{fig:nogo}
\end{figure}

By the middle of the second semester, Stella's interest in entering the quantum industry had waned even more. In response to a photovoice prompt reading, ``Take a photo representing your level of interest at the moment about pursuing a career in quantum industry,'' Stella posted the photo shown in Fig.~\ref{fig:nogo} with the caption, ``as of right now, it's a `no go.' ''

In the focus group in the middle of the spring semester, Stella spoke about this photo, indicating that she took it with intention. She said that she waited at this intersection for the light to turn red so that she could take the photo. This indicates that Stella was looking to represent her feelings about not wanting to go into quantum industry photographically, rather than identifying feelings to go along with a photo she took by chance. The certainty implied in this photo and its caption, along with Stella's discussion of it, represent a shift in Stella's perspective since the end of the first semester. 

Like Jasper, Stella also felt discouraged from joining the quantum industry due to its lack of representation. When asked by the interviewer ``have you ever felt uncomfortable or excluded from quantum industry or spaces that would help you prepare for it?'' Stella replied, 

\begin{quote}
    ``Yes. Yeah, just general stuff with the sexism, which is very not fun to deal with. So in that way, I felt excluded. But that's pretty much the only- only way I felt excluded from industry. And it wasn't even the industry's fault. It was just my observations, like, oh, there's not a lot of women here. And then in the project, I feel like it has amplified, sort of, that feeling of like, oh, this is a little uncomfortable to, like, I'm not really feeling like being here. And I don't think it's that quantum industry or any industry in STEM doesn't want to hire women. They definitely do. It's just that I don't know if I'd have a good time.''
\end{quote}
This quote suggests that the lack of representation of women that Stella observed in the industry, as well as her experiences in the course, contributed to her lack of desire to join it. 

Not only was Stella discouraged by seeing a lack of diversity in the industry, but she was also discouraged by her experience of what she considered to be microaggressions within her team. Her experience with this group dynamic suggested to her that the quantum industry might not be able to support her as a member of a marginalized group, causing her interest to decline. Later in the interview, in response to a prompt about whether she was interested in entering the quantum industry, she continued, 

\begin{quote}
    ``Maybe one day, so currently no. Well, I wouldn't say- not like a hard no, like a soft no. Like, probably not. Just because of the reasons we've talked about [regarding sexism and microaggressions]. And, like, yeah, I'm not completely closed off to the idea of it. But I don't think that's what I'm going to go for right now. Like when I- after I graduate.
\end{quote} 

She also said in the same interview, 

\begin{quote}
     ``I know there's HR departments, I'm gonna go [somewhere] with an HR department 100\%. So probably not a startup. But a lot of quantum industry is startup. So I'm not sure if I would trust a small company to be able to handle issues like [microaggressions and sexism]. Which is the only reason that I'm- feel like I'm not really interested right now.''

\end{quote}

Stella once again indicated that the lack of diversity that she saw in the industry was discouraging her from wanting to participate. As mentioned above, addressing the lack of diversity within the field may be important to supporting students like Stella. Furthermore, helping students understand microaggressions and the harm that they can cause may help prevent students from committing microaggressions and prevent similar issues in future courses. Finally, it is important that companies within the quantum industry design mechanisms for handling issues like microaggressions and that these mechanisms are communicated to interested students. 

\subsubsection{\label{Nina}Nina}

Nina also identified herself to us as a white woman. While Nina was interested in participating in the quantum industry, she ultimately had some trepidation about it and ended up considering other options alongside the quantum industry. Nina participated in an interview with us at the end of the first semester and the end of the second semester, along with her responses to reflection questions and photovoice prompts throughout the year. 

\begin{figure}[t!]
    \includegraphics[width=0.4\textwidth]{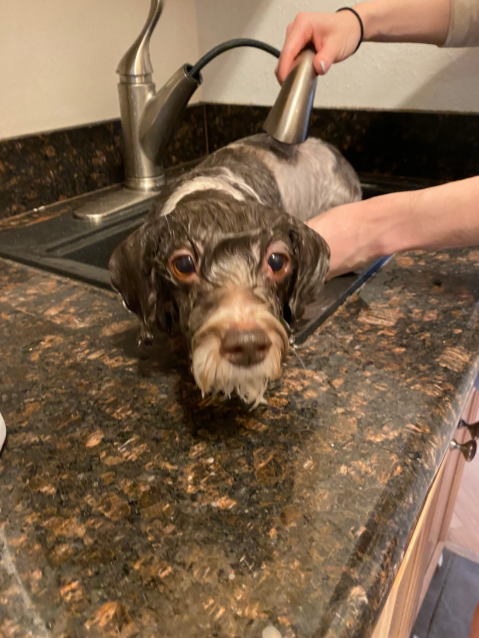}
    \caption{A photo of a wet brown and white dog being washed in a sink in response to the photovoice prompt ``Take a photo representing your level of interest at the moment about pursuing a career in quantum industry.'' This photo was captioned: ``I'm a little unsure like my roommates dog getting a bath.''}
    \label{fig:dog}
\end{figure}

Nina's uncertainty about the quantum industry was captured in her response to a photovoice prompt asking, ``Take a photo representing your level of interest at the moment about pursuing a career in quantum industry,'' Nina responded with the photo of a dog in Fig.~\ref{fig:dog}. Her caption read, ``I'm a little unsure like my roommates dog getting a bath.''
This emphasized the fact that Nina did not have particularly strong feelings one way or another about the quantum industry, and she was experiencing some hesitancy about entering the industry. 

While she did not elaborate on her reasons for being hesitant in response to this photovoice prompt, she gave some additional insight into this in her interview at the end of the academic year. When asked what her plans were for after graduation, Nina replied,``I'm still applying to jobs... so some sort of industry work.''

The interviewer then asked if she was talking specifically about quantum industry and Nina elaborated,

\begin{quote}
    ``No, not specifically. Maybe. But some things I'm looking at and then, like, software things as well. I'm kind of interested in pursuing a job anywhere, like, technical or physics-y just because I don't really know what I want to do and am in the place where I just want to try something, try a job and see what aspects of it I like.''
\end{quote}

Nina's response to this series of interview questions suggests that her lack of strong feelings about the quantum industry came from not truly understanding what aspects of a job she would enjoy and just wanting to ``try something'' in order to understand that more deeply. A class such as Q-Forge may be able to support students like Nina by having an additional focus on what entry-level jobs in the quantum industry look like on a day-to-day basis, so that students can better understand whether they think they would enjoy that type of work.

Nina was also concerned about the lack of diversity she saw within the quantum industry. She said during the mid-course interview,

\begin{quote}
    ``Something that, like, a lot of the team has been talking about is, like, when we go to visit [our industry partner] we notice, like, `oh, it's a really cool company, with really cool ideas. But everyone working there is, like, young white men' and kind of realizing, like, company cultures and, like, environments, that I do or don't want to work in is a huge- like, there is not a single woman there. Um, which I'm sure isn’t uncommon in quantum industry.
\end{quote}

As mentioned above, students are quite aware of the lack of diversity within the quantum industry (and physics more broadly). While these issues of representation must be resolved on a larger scale, classes like Q-Forge have a responsibility to help support students as they notice the lack of diversity and representation within the industry. More research is needed to understand how best to do this. Some potential options include facilitating conversations about diversity in the industry among students and helping them network with marginalized individuals in the industry.

There were also, however, many reasons why Nina was interested in the quantum industry. First, she discussed working on the industry project as motivating her to want to participate in quantum industry. In response to a reflection question at the end of the second semester reading ``how has working on this industry project impacted your interest in pursuing a career in industry?'' Nina wrote, ``It's helped me see how exciting industry is and how answers are mostly unknown.'' 

Nina also brought up the excitement and novelty of the industry in her end-of-year interview. In response to a question about whether or not she felt like she would belong in the quantum industry, Nina said, 

\begin{quote}
    ``I think I would, like, belong in quantum industry based on the, like, curiosity and kind of excitement for where it can take us.''
\end{quote}

Similarly to Jasper, Nina felt that she was excited about the quantum industry and saw that others who were participating in the industry were excited about the potential discoveries they could make. As mentioned above, allowing students to see the excitement of the industry and the possibilities for ``where it can take us'' partially through courses such as Q-Forge may be an important piece of motivating students to be interested in and excited about the industry. 

Finally, when asked in the interview at the end of the first semester about how the course influenced her interest in quantum industry, Nina discussed the visits the Q-Forge students had with individuals representing local quantum companies:

\begin{quote}
    ``I think, like, going to visit actual companies... otherwise I wouldn't have, kind of, a visual for what those work environments look like, or what to look for in quantum industry.''
\end{quote}

While these visits did not make Nina feel particularly strongly one way or another about wanting to enter the quantum industry, she did feel that they were important to helping her begin to understand what the industry looks like in action. The value of this type of experience, even if it discourages some students from entering the industry, cannot be discounted; helping students learn that they are less interested in the industry than they originally thought is also valuable. 

\subsubsection{\label{Owen}Owen}

Owen identified himself to us as a white man. He had generally high levels of interest in participating in the quantum industry throughout the course, and was interested in any type of job within the industry. Owen participated in interviews with us at the end of the first semester and the end of the second semester, and he participated in reflection questions and photovoice throughout.

Owen often reaffirmed his generally high levels of interest in quantum industry in response to reflection questions and interview questions. For instance, when faced with a reflection question at the end of the first semester that asked ``how has working on this industry project impacted your interest in pursuing a career in quantum industry?'' Owen replied simply, ``I was interested before and am still interested now.'' This suggests that, while the project had little impact on his interest, this was likely because he was already quite interested before beginning the project. 

In his interview at the end of the first semester, Owen clarified that he did not feel particularly tied to a specific type of job within the industry. Owen said, ``I think if I had a job in the quantum industry I would be happy,'' indicating that his happiness was not dependent on a specific type of role or specific type of company. This interest persisted throughout Owen's responses for most of the year. 

Owen suggested that participating in the industry project ultimately did support his interest in going into quantum industry. In his mid-course interview, in response to a question about how the course influenced his interest in the quantum industry, he said that he thought the course had only strengthened his interest. When asked in what ways, he replied,

\begin{quote}
    ``You know, with just how important our project is, I think, to [the industry partner], you know, they really gave us- they give us a problem that's really integral to their product. It seems like we're actually doing something that would be really helpful. That stands to be pretty unique in the- in the industry, so, yeah, I just think it's- how important the project is to- to [the industry partner].''
\end{quote}

This suggests that, once again, the authenticity of the project that the students worked on was a driving factor behind this student's interest in the industry. Allowing students to participate in a project that has objective value to someone outside of the university may be an asset of courses like Q-Forge in helping to attract new talent to the industry, as well as a variety of other reasons including improving student motivation. 

Despite Owen's continued interest in the quantum industry throughout the course, at his final interview he expressed some hesitation about joining it, although his hesitation was not due to any aspect of the industry directly. He was particularly interested in working on some of his own ideas rather than entering the workforce at all directly after graduation. Particularly, he said,

\begin{quote}
    ``I definitely will throw some applications around, I will absolutely throw some apps around. With what fervor I will pursue these opportunities, I don't know yet. But I'll throw my hat in the ring for sure.''
\end{quote}

Additionally, he said, 

\begin{quote}
    ``I got some of my own stuff I want to work on. I've had some- I've had some ideas brewing of projects that I want to take on that have nothing to do with quantum computing. And I want to take a crack at those.''
\end{quote}

This indicates that Owen was still interested in becoming a part of the quantum industry to some degree at the end of the course, but that he had his own personal reasons for perhaps not being as interested as he was at the beginning of the year. This goes to show that students' choices to enter the quantum industry or not are deeply personal and, in some cases, may be influenced by things that a course like Q-Forge cannot control. 

\section {\label{Conclusions}Conclusions}
\subsection{Summary}

Throughout this work, we answered three research questions, the answers to which are summarized below.

\textbf{RQ 1} asked: What do students think that the quantum industry is? We began to answer this research question by identifying that these students knew very little about the quantum industry at the beginning of the year. It is important to note that even upper-division students, many of whom had already taken one or two semesters of quantum mechanics, may still know very little about the quantum industry at the start of a course like this. Furthermore, the students we studied were students who explicitly chose to take a course focused on the quantum industry, and their background knowledge was still limited. 

Over the course of the year, the students learned that the quantum industry is broader than they thought and does not focus strictly on quantum computing. This understanding of the breadth of the industry was a benefit of the students working on the type of project they worked on, as they were able to understand more deeply the side of the industry that supports other quantum endeavors. 

The students also learned that the quantum industry lacks diversity in a similar way to physics more broadly. This aspect of the quantum industry was not a negative just for students from marginalized groups. The fact that students were observing this about the industry highlights the importance of increasing diversity in the quantum workforce. While the diversity of the industry cannot be expected to change overnight, it may be important to involve students in discussions about this part of the industry and help them express their concerns, as well as involve them in eliciting change.

Despite the students noticing this negative fact about the industry, they overall felt that the quantum industry was exciting. They brought up that it was involved in creating exciting change in technologies, and also that it was filled with people who are excited to make those breakthroughs and change the landscape of quantum technologies. 

Students learned about the quantum industry in many ways, but one important way was through the visits that they had with local quantum companies. These visits could be an important aspect of a primarily capstone-oriented course to help students learn about the industry outside of their specific project, or they could be an addition to a non-capstone-style course or degree program seeking to incorporate some aspect of education about the quantum industry. 

~\textbf{RQ 2} asked:  Do students think they can be successful in the quantum industry? One of the most common things brought up by students was that they felt like they needed to have a graduate degree in order to get a job in the industry. While it does seem that many jobs in the U.S. and Europe require Master's degrees or Ph.D.s, more research is needed to understand how to direct these undergraduate students towards roles that would be appropriate for their level of education and better understand what the industry is looking for from employees with Bachelor's degrees.

Students both felt like they had some of the skills required to be successful in quantum industry and felt that they were missing others. Helping students acquire and understand that they have the relevant skills for pursuing a job in their industry is an important role for a course like Q-Forge. Designers of similar courses should consider methods of helping students understand what skills companies expect them to have and to assess whether or not they have those skills. 

Finally, students felt that their involvement with the industry project helped them feel as though they could be successful in quantum industry. This authentic industry experience was extremely valuable to these students, and should be considered an important aspect of a course attempting to train a quantum workforce. 

~\textbf{RQ 3} asked: Do students want to participate in the quantum industry? The answer to this question is complicated and deeply personal. Most of the students in this course entered with an interest in going into quantum industry, and some discovered that there were other paths that were more appealing to them. Regardless, it is important to note that we may not be able to, or want to, encourage all students towards participation in the quantum industry; giving them the information required to have agency over their decision to participate in the industry at any given time is an equally valuable role of a course like Q-Forge.

\subsection{Recommendations for Future Instructors and Course Designers}

Based on the experiences of these five students in Q-Forge, we propose several recommendations for future quantum industry courses and degree programs. While these recommendations are based on the perceptions of a small number of students, we feel that they hold value for influencing decisions about future quantum industry courses and programs, as they provide a student perspective to discussions that are often dominated by faculty and other non-student stakeholders. 

\emph{It is important to note the value of the authentic industry project}. Many of these students brought up the role that their industry project played in their knowledge about, ability to succeed in, and interest in the quantum industry. As one of these students mentioned, perhaps the most important job preparation one can provide for students is relevant experience in the area of interest, and the ability of an authentic industry project to introduce students to what actually working in the quantum industry is like cannot be downplayed. Authentic industry projects may also be useful to students entering other types of industry, not solely quantum. 

\emph{Expose students to the breadth of the industry} in courses like Q-Forge. Students in Q-Forge identified working on a project outside of quantum computing as a key way that they learned about the breadth of the industry, but this can also be done through activities like visits to or from companies that do work in varying sub-areas of the industry.

\emph{Talk about diversity} in courses like Q-Forge. Students notice the lack of diversity in what they imagine might be their future workplace and it is likely important that they be able to have conversations about how that affects them and their classmates. 

\emph{Recognize that these are undergraduate students} and help them network with, and understand the experiences of, individuals with similar backgrounds to them. It is important that these students understand what types of jobs they might be able to get without further education, what role they might play in a company, and what skills are required to get them there. Other individuals with similar educational backgrounds who are already participating in the industry can help them understand these details and nuances. 

\emph{Help students develop relevant skills}, both for a first job and a future career, that are desired by the quantum industry, as well as better understand what these skills are. This will help students assess their ability to be successful in the industry and ensure that they are prepared to enter the industry after graduation. It is important, however, that this point should be achieved in conjunction with an authentic industry project, as developing skills without the context of industry work may fail to give students the direct industry experience they need to be successful. 

\emph{Help students develop knowledge about the quantum industry.} Since students are not always deeply aware of what the quantum industry is and does, it is important that instructors structure courses like Q-Forge to provide students with an introduction to the industry rather than assuming that they have gathered this information elsewhere. This includes introducing students to the types of work that are typically done by individuals with Bachelor's degrees within the industry. This can be done through in person or virtual visits with companies in the industry, for example. 

\subsection{Future Work}

As this work begins to help us understand the nature and value of courses like Q-Forge, we plan to continue investigating the role of this course in developing a quantum workforce, and encourage others to investigate questions about how undergraduate students might participate in the quantum industry and how to best prepare them for those opportunities. 

We plan to follow up with these five students a year after their graduation in order to discover what types of jobs they have worked in over the past year, whether they have actually become involved in the quantum industry and, if so, what that experience has been like for them. This will help us understand the longer term value of Q-Forge and whether it has actually prepared students to enter the quantum industry. 

We are also continuing to investigate the outcomes of this course in its second iteration during the 2023-2024 academic year. Changes were made to this course after feedback from the authors at the end of the 2022-2023 academic year, meaning that the second iteration of this course may well be different than the first. Furthermore, the number of students enrolled in the course has grown, as well as the number of projects offered to students. We hope to continue to be able to provide information to the community about the value of these types of courses and how they can be executed more effectively. 

Additional work is needed to understand the quantum landscape as it grows and changes. In order to better understand whether teaching courses like Q-Forge is worth the required resources, we must better understand the availability of jobs in the quantum industry for workers with physics Bachelor's degrees and no additional work experience, as well as the skill set desired by the industry for these positions. Equipped with this information, we will be better able to address the industry's need for more qualified workers. 

Finally, as courses like Q-Forge proliferate across the U.S. and around the world, we encourage others to study student perceptions of these hands-on, authentic courses and what they are achieving with respect to preparing a quantum-ready workforce. For instance, it may be important to understand how participation in a capstone course like Q-Forge compares to participation in a quantum research lab or formal internship. Furthermore, the efficacy of these courses in improving diversity of the quantum industry should be studied. Through these efforts, we can better understand how to improve these courses for the future and increase our efficacy at preparing students to become a part of this important industry. 

\begin{acknowledgments}
We would like to express our deepest gratitude to the students of Q-Forge 2022-2023 for their willingness to work with us on this project. Their insight is incredibly valuable to the community of quantum educators, researchers, and industry members. We would also like to thank the Q-Forge instructor, Charles Rogers, and course designer, Mike Bennett, for graciously accepting us into their classroom to conduct this research. Finally, we would like to thank
the PER group at the University of Colorado Boulder
for their feedback on this work. This work is supported
by the National Science Foundation QLCI Award OMA
2016244. 
\end{acknowledgments}

\bibliography{QuantumIndustry}

\clearpage
\includepdf[pages={{},1,{},2,{},3,{},4}]{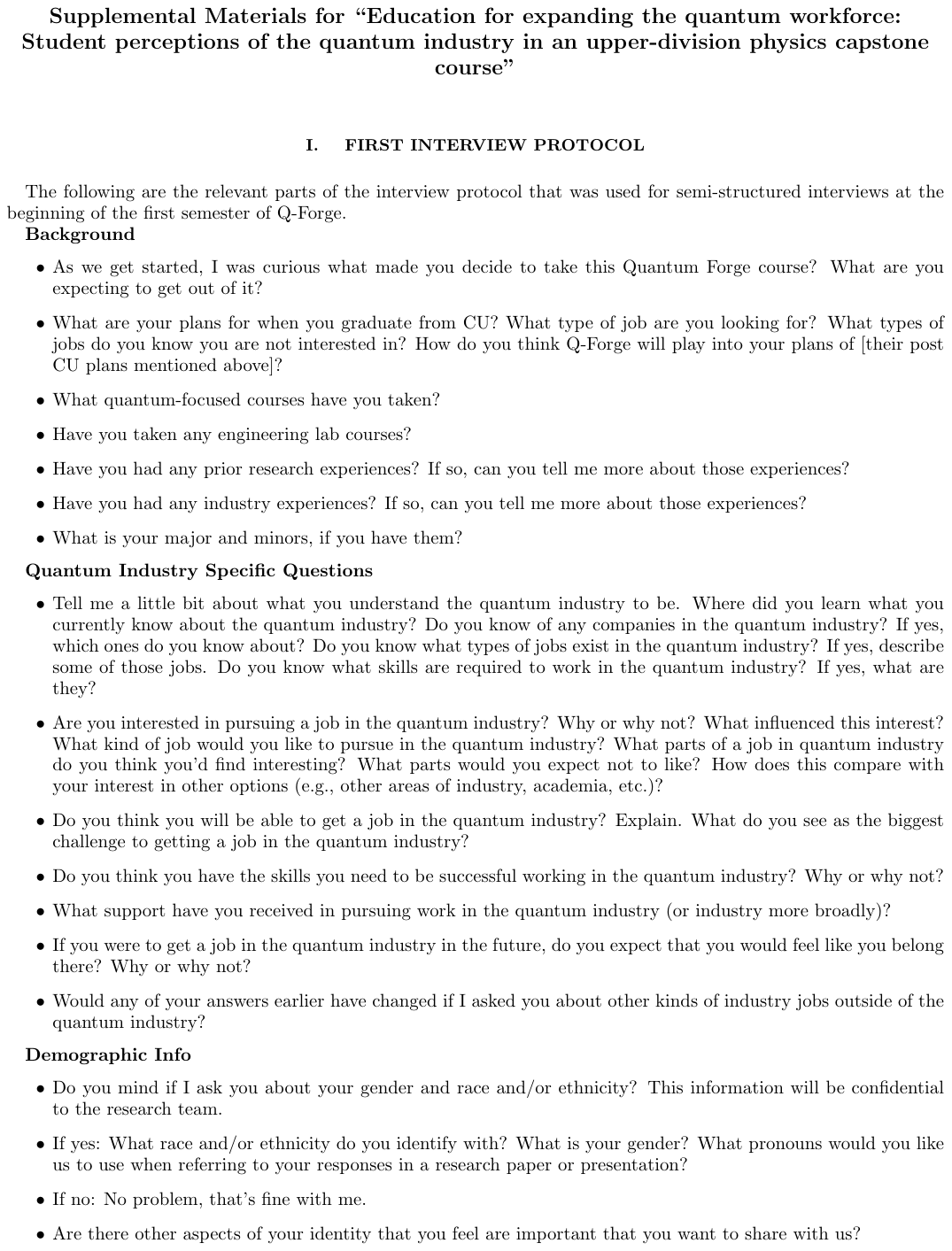}

\end{document}


\title{Supplemental Materials for ``Education for expanding the quantum workforce: Student perceptions of the quantum industry in an upper-division physics capstone course''}

\maketitle

\section{\label{InterviewP1} First interview protocol}
The following are the relevant parts of the interview protocol that was used for semi-structured interviews at the beginning of the first semester of Q-Forge. 

\textbf{Background}
\begin{itemize}
    \item As we get started, I was curious what made you decide to take this Quantum Forge course? What are you expecting to get out of it?
    \item What are your plans for when you graduate from CU? What type of job are you looking for? What types of jobs do you know you are not interested in? How do you think Q-Forge will play into your plans of [their post CU plans mentioned above]?
    \item What quantum-focused courses have you taken? 
    \item Have you taken any engineering lab courses?
    \item Have you had any prior research experiences? If so, can you tell me more about those experiences? 
    \item Have you had any industry experiences? If so, can you tell me more about those experiences?
    \item What is your major and minors, if you have them?
\end{itemize}

\textbf{Quantum Industry Specific Questions}
\begin{itemize}
    \item Tell me a little bit about what you understand the quantum industry to be. Where did you learn what you currently know about the quantum industry? Do you know of any companies in the quantum industry? If yes, which ones do you know about? Do you know what types of jobs exist in the quantum industry? If yes, describe some of those jobs. Do you know what skills are required to work in the quantum industry? If yes, what are they?
    \item Are you interested in pursuing a job in the quantum industry? Why or why not? What influenced this interest? What kind of job would you like to pursue in the quantum industry? What parts of a job in quantum industry do you think you’d find interesting? What parts would you expect not to like? How does this compare with your interest in other options (e.g., other areas of industry, academia, etc.)?
    \item Do you think you will be able to get a job in the quantum industry? Explain. What do you see as the biggest challenge to getting a job in the quantum industry?
    \item Do you think you have the skills you need to be successful working in the quantum industry? Why or why not? 
    \item What support have you received in pursuing work in the quantum industry (or industry more broadly)? 
    \item If you were to get a job in the quantum industry in the future, do you expect that you would feel like you belong there? Why or why not?
    \item Would any of your answers earlier have changed if I asked you about other kinds of industry jobs outside of the quantum industry?
\end{itemize}

\textbf{Demographic Info}
\begin{itemize}
    \item Do you mind if I ask you about your gender and race and/or ethnicity? This information will be confidential to the research team.
    \item If yes: What race and/or ethnicity do you identify with?  What is your gender? What pronouns would you like us to use when referring to your responses in a research paper or presentation?
    \item If no: No problem, that’s fine with me.
    \item Are there other aspects of your identity that you feel are important that you want to share with us?
\end{itemize}

\section{\label{InterviewP2} Second and third interview protocol}

The following are the relevant parts of the interview protocol  used for semi-structured interviews at the end of the first and second semesters of Q-Forge. If students did not participate in a first interview, demographic questions were also asked here. 

\textbf{Background}
\begin{itemize}
    \item Have your plans for a job/career after graduation changed since the previous interview?
\end{itemize}

\textbf{Quantum Industry Specific Questions}
\begin{itemize}
    \item Tell me a little bit about what you now understand the quantum industry to be after taking this course. Do you feel like you’re more familiar with what companies exist in the quantum industry after taking this course? How did you learn about these companies? 
\item Do you feel like you have more knowledge now of what types of jobs exist in the quantum industry than you did at the beginning of the semester? Describe some of those jobs. Of those, which would you consider/be interested in? Of those, which are you not interested in?
\item Do you have more knowledge of what skills are required to work in the quantum industry? How did your knowledge of these skills change? 
\item If you pursued a job in the quantum industry, what do you imagine the work would be like? What parts do you expect you’d find interesting? What parts would you expect not to like?
\item Are you currently interested in pursuing a job in the quantum industry? Why or why not? What parts of the course influenced this interest? How does this compare with your interest in other options (e.g., other areas of industry, academia, etc.)? How has this changed since the beginning of the semester?
\item How confident are you currently in your ability to get a job in the quantum industry? Explain. What do you see as the biggest challenge to getting a job in the quantum industry? How has this changed since the beginning of the semester?
\item Do you think you currently have the skills you need to be successful working in the quantum industry? Why or why not? 
\item What support have you received this semester in pursuing work in the quantum industry (or industry more broadly)? 
How has this changed since the beginning of the semester?
What support do you feel like Q-Forge gave you? ‘
\item Have you ever felt uncomfortable or excluded from the quantum industry or spaces or activities that would help you prepare for it this semester? In what ways? What would you need to feel like you belong in the quantum industry? Do you think you are a good fit for the quantum industry? How has this changed since the beginning of the semester?
\end{itemize}

\section{\label{RQList} Reflection Question Prompts}
The following are the reflection questions related to student perceptions of the quantum industry that were asked of students throughout the course of the year. The prompts followed by an asterisk cycled through in the second semester every three weeks (3-4 times total) after the students settled into their routine of working on the industry project.  

\textbf{First Semester}
\begin{itemize}
    \item Was there anything surprising you learned during the discussion about intellectual property and non-disclosure agreements? Briefly describe what it was, why you found it surprising, and how it compares with the way ideas were discussed in your prior research and/or industry experiences, if you have any.
    \item After doing the essential skills modules [project management modules] this week, how comfortable do you feel using industry-standard terminology to communicate what are the necessary inputs and potential outputs of your project?
    \item At the moment, how confident are you about your ability to contribute to the [industry] project and why?
    \item What did you learn from the field trip [to industry sponsor] both with respect to the quantum industry and with respect to your project? 
    \item After starting to work on the project, has your interest in pursuing a job in the quantum industry increased, decreased, or remained the same? Please explain.
    \item How have your interactions with the industry partners been? What has gone well? What could be done better?
    \item Describe one thing you did this week that you think is authentic to a real industry experience.
    \item Describe one thing you did this week related to Q-Forge that you do not think is part of an authentic industry experience.
    \item To what extent do you feel that the support you've received from the course instructors and [the industry partner] has been adequate to ensure that you make progress on the project? Please explain.
    \item How are you feeling about your team's ability to make progress on the project given the updated expectations and project goals? 
    \item You have now visited or had visitors from four local quantum companies. What have you gotten out of your interactions with various quantum industry companies aside from [the industry partner]?
    \item How has working on this industry project impacted your interest in pursuing a career in industry?
    \item After participating in this industry project, how confident do you feel in your ability to successfully work in industry? Please explain.
\end{itemize}

\textbf{Second Semester}
\begin{itemize}
    \item How did you feel about your team's communication with the industry sponsor this week?* 
    \item Did you do anything in the past week that helped prepare you for a job in quantum industry? If so, what was it? If not, was there anything else you did for Q-Forge that you found valuable?*
    \item Did you have any experiences in the course this week that changed your level of interest in entering the quantum industry? Why or why not?*
    \item What skills do you feel like you didn't gain from this course that are important to being successful in quantum industry?
    \item Are you planning to enter the quantum industry after graduating? Why or why not?
\end{itemize}

\section{\label{PVPrompts} Photovoice Prompts}
The following are the photovoice prompts related to students perceptions of the quantum industry that were used throughout the Q-Forge course. The prompts followed by an asterisk were cycled through every three weeks (3-4 times total) in the second semester after the students settled into working on the industry project:

\textbf{First Semester}
\begin{itemize}
    \item Take a photo that represents the environment/culture of a company in the quantum industry.
    \item Take a photo representing how you feel about the project right now.
    \item Take a photo representing what you find most interesting at the moment about pursuing a career in quantum industry.
    \item Take a photo of something with respect to the [industry] project that you are proud of.
\end{itemize}

\textbf{Second Semester}
\begin{itemize}
    \item Take a photo representing your level of interest  at the moment about pursuing a career in quantum industry.*
\end{itemize}

\section{\label{FGProtocol} Photovoice Focus Group Protocol}
The following questions were asked of the students with respect to each category of photos during each of the focus groups: 
\begin{itemize}
    \item What themes do you notice in these photos?
    \item How do these themes represent your collective experience with quantum industry/the course as a whole? 
    \item Do any of the comments other people made about your photos resonate with you? Were there any comments that surprised you?
    \item After hearing from the photographers, are there any changes you want to make to the list of themes you noticed in the photos? 
\end{itemize}